\documentclass[12pt]{article}
\usepackage{times}
\usepackage{geometry}
\usepackage[utf8]{inputenc}
\usepackage{enumitem,amssymb}
\usepackage{ragged2e}
\usepackage{hyperref}
\newlist{thematic}{itemize}{8}
\setlist[thematic]{label=$\square$}
\usepackage{pifont}
\usepackage{natbib}
\newcommand{\cmark}{\ding{51}}%
\newcommand{\done}{\rlap{$\square$}{\raisebox{2pt}{\large\hspace{1pt}\cmark}}%
\hspace{-2.5pt}}

\begin{document}
{
\raggedright
\huge
Astro2020 Science White Paper \linebreak

Cyberinfrastructure Requirements to Enhance Multi-messenger Astrophysics\linebreak
\normalsize

\noindent \textbf{Thematic Areas:} \hspace*{60pt} $\square$ Planetary Systems \hspace*{10pt} $\square$ Star and Planet Formation \hspace*{20pt}\linebreak
$\square$ Formation and Evolution of Compact Objects \hspace*{31pt} $\square$ Cosmology and Fundamental Physics \linebreak
  $\square$  Stars and Stellar Evolution \hspace*{1pt} $\square$ Resolved Stellar Populations and their Environments \hspace*{40pt} \linebreak
  $\square$    Galaxy Evolution   \hspace*{45pt} $\done$             Multi-Messenger Astronomy and Astrophysics \hspace*{65pt} \linebreak
  
\textbf{Principal Author:}

Name:	Philip Chang
 \linebreak						
Institution:  University of Wisconsin - Milwaukee
 \linebreak
Email: chang65@uwm.edu
 \linebreak
 
\textbf{Co-authors:} 
%
%
Gabrielle Allen$^{1}$,
 Warren Anderson$^{2}$,
 Federica B. Bianco$^{3,4}$,
 Joshua S. Bloom$^{5,6}$,
 Patrick R. Brady$^{2}$,
 Adam Brazier$^{7}$,
 S. Bradley Cenko$^{8}$,
 Sean M. Couch$^{9}$,
 Tyce DeYoung$^{9}$,
 Ewa Deelman$^{10}$,
 Zachariah B Etienne$^{11,12}$,
 Ryan J. Foley$^{13}$,
 Derek B Fox$^{14}$,
 V. Zach Golkhou$^{15}$,
 Darren R Grant$^{9}$,
 Chad Hanna$^{14,16,17}$,
 Kelly Holley-Bockelmann$^{18,19}$,
 D. Andrew Howell$^{20,21}$,
 E. A. Huerta$^{22,1}$,
 Margaret W.G. Johnson$^{22,1}$,
 Mario Juric$^{15}$,
 David L. Kaplan$^{2}$,
 Daniel S. Katz$^{1}$,
 Azadeh Keivani$^{23}$,
 Wolfgang Kerzendorf$^{24}$,
 Claudio Kopper$^{9}$,
 Michael T. Lam$^{11}$,
 Luis Lehner$^{25}$,
 Zsuzsa Marka$^{23}$,
 Szabolcs Marka$^{23}$,
 Jarek Nabrzyski$^{26}$,
 Gautham Narayan$^{27}$,
 Brian W. O'Shea$^{9}$,
 Donald Petravick$^{22,1}$,
 Rob Quick$^{28}$,
 Rachel A. Street$^{20}$,
 Ignacio Taboada$^{29}$,
 Frank Timmes$^{30}$,
 Matthew J. Turk$^{1}$,
 Amanda Weltman$^{31}$,
 Zhao Zhang$^{32}$

\medskip
$^{1}$University of Illinois at Urbana-Champaign,
$^{2}$University of Wisconsin - Milwaukee,
$^{3}$University of Delaware,
$^{4}$Center for Urban Science and Progress, NYU,
$^{5}$University of California, Berkeley,
$^{6}$Lawrence Berkeley National Laboratory,
$^{7}$Cornell University,
$^{8}$NASA GSFC,
$^{9}$Michigan State University,
$^{10}$University of Southern California,
$^{11}$West Virginia University,
$^{12}$Center for Gravitational Waves and Cosmology, WVU,
$^{13}$University of California, Santa Cruz,
$^{14}$The Pennsylvania State University,
$^{15}$University of Washington,
$^{16}$Institute for Gravitation and the Cosmos, PSU,
$^{17}$Institute for CyberScience, PSU,
$^{18}$Vanderbilt University,
$^{19}$Fisk University,
$^{20}$Las Cumbres Observatory,
$^{21}$University of California, Santa Barbara,
$^{22}$NCSA,
$^{23}$Columbia University in the City of New York,
$^{24}$New York University,
$^{25}$Perimeter Institute for Theoretical Physics,
$^{26}$University of Notre Dame,
$^{27}$Space Telescope Science Institute,
$^{28}$Indiana University,
$^{29}$Georgia Institute of Technology,
$^{30}$Arizona State University,
$^{31}$University of Cape Town,
$^{32}$Texas Advanced Computing Center

\pagebreak

\noindent\textbf{\Large Abstract}
\smallskip

The identification of the electromagnetic counterpart of the gravitational wave event, GW170817, and discovery of neutrinos and gamma-rays from TXS\,0506+056 heralded the new era of multi-messenger astrophysics. As the number of multi-messenger events rapidly grow over the next decade, the cyberinfrastructure requirements to handle the increase in data rates, data volume, need for event follow up, and analysis across the different messengers will also explosively grow.  The cyberinfrastructure requirements to enhance multi-messenger astrophysics will both be a major challenge and opportunity for astronomers, physicists, computer scientists and cyberinfrastructure specialists.  Here we outline some of these requirements and argue for a distributed cyberinfrastructure institute for multi-messenger astrophysics to meet these challenges.   

\bigskip

\noindent {\Large \bf Scientific Motivation}
\smallskip

With the identification of the electromagnetic counterparts of both gravitational wave and high-energy neutrino events over the past two years, the new era of multi-messenger astrophysics (MMA) has begun. The observations of a binary neutron star merger in gravitational waves
(GW/GRB\,170817A; \citealt{2017PhRvL.119p1101A}), along with subsequent worldwide electromagnetic discoveries \citep{2017ApJ...848L..12A}, and of neutrinos and gamma-rays from TXS\,0506+056 \citep{2018Sci...361.1378I} demonstrate both the promise and the challenges of MMA. These events will allow us to answer questions such as the origin of the heaviest elements, the equation of state of extremely dense matter in mergers of compact systems, the demographics of the time-domain sky, and the rate of the expansion of the Universe.

The next decade will see dramatic improvements in MMA capabilities.  Upgrades of existing and deployment of new GW detectors such as LIGO \citep{2015CQGra..32k5012A}, VIRGO \citep{2015JPhCS.610a2014A}, and KAGRA \citep{2019NatAs...3...35K} will enable the identification of dozens of neutron-star merger events
per year. IceCube-Gen2 \citep{2014arXiv1412.5106I,2017JInst..12P3012A} and other enhanced neutrino and cosmic ray observatories will continue to survey the sky for
high-energy cosmic particles. NANOGrav \citep{2013CQGra..30v4008M} and other Pulsar Timing Arrays (PTAs) will probe the gravitational wave spectrum at nanohertz frequencies. the Large Synoptic Survey Telescope (LSST; \citealt{2008arXiv0805.2366I}) and the Zwicky Transient Facility \citep{2019arXiv190201945G} will
survey the optical sky with unprecedented speed and depth. Upcoming radio facilities will collect terabytes of data with unprecedented resolution. This range of astronomical observatories will gather data on candidate multi-messenger sources, source populations, and host galaxies throughout the visible Universe. A comprehensive cyberinfrastructure for MMA will coordinate the efforts of these facilities, while providing valuable benefits for all areas of astrophysics that are part of their core scientific missions.  

The explosive growth in the scale and rates of data acquisition, the importance of event follow-up in MMA, and the challenge of analysis over different messengers drive the need for cyberinfrastructure upgrades that will require the sustained collective efforts of astrophysics facilities and a broad-ranging group of physicists, astronomers, computer scientists, data science specialists, and software/infrastructure engineers. {\bf Cyberinfrastructure will need to be designed, built, and operated in order to collect, process, analyze, synthesize, and enable interpretation time-critical and archival data from observing facilities across the multi-messenger spectra}.  Robust tools for data sharing, collaboration, joint analysis, as well as training and education will also need to be developed.  
Investments will enable sustainable and efficient operations leading to more compelling discoveries, scientific outcomes, public releases of data and supporting analysis codes, cross-disciplinary research, cross-collaboration work, training and education of a diverse STEM workforce, and building a community-driven integrated software infrastructure. 

\bigskip
\noindent {\Large \bf Cyberinstructure For Multi-Messenger Astrophysics}
\smallskip

By its nature, MMA requires diverse scientific teams to bring together their observational
resources, data, analysis and modeling tools, and expertise to ensure the maximum scientific return.  With a diversity of different instruments, systematic effects,
data formats, analysis protocols, collaboration styles, proprietary and other restrictions, and scientific cultures come a myriad of challenges. Working together effectively in this heterogeneous and international scientific federation requires a common, scalable and extensible
framework for cooperation, involving communication and engagement among instruments, collaborations, projects, individual scientists, and citizen scientists.
These larger groups require the ability to form flexible teams on short time scales, to share data, codes and other digital objects in real time, and to self-organize into spontaneous collaborations of varying scope.  Additionally, the differing hardware infrastructures and computational resources may also need to be shared and mutually understood.
On the organizational level, teams may need to share proprietary data and information governed by memoranda of understanding (MoUs). This may be facilitated by providing a means to hold data in ``escrow'' and give policy-driven access permissions to that data. 

Although important, these are merely the ``known unknowns'' of cyberinfrastructure challenges for MMA.  To map out the entire problem space, the Scalable Cyberinfrastructure for Multi-messenger Astrophysics (SCiMMA)\footnote{\url{https://scimma.org}} team has been meeting since August of 2018 to define the requirement and challenges in the areas of cyberinstructure that can provide common, value-added services for scientists across the multi-messenger spectrum, to discuss with major MMA stakeholders--such as the Laser Interferometer Gravitational-Wave Observatory, IceCube,  LSST, Laser Interferometer Space Antenna, the North American Nanohertz Observatory for Gravitational Waves (NANOGrav), National Optical Astronomy Observatory (NOAO), Cherenkov Telescope Array (CTA; \citealt{2011ExA....32..193A}), NASA and others---their major cyberinfrastructure requirements, and to formulate a plan to meet these challenges. 

In January of 2019, a number of astrophysicists, computer scientists, and cyberinfrastructure specialists met at Columbia University to discuss the cyberinfrastructure requirements and challenges of MMA.  The two day workshop revealed a number of important open questions and areas of cross disciplinary research and interest.  The workshop concluded that: 
\begin{itemize}
  \item \textbf{Software: } There is a need for community-oriented and -driven software that can 1) access MMA data, 2) analyze multiple types of MMA data, 3) model MMA sources, 4) minimize and facilitate new development by building on and contributing to common libraries and frameworks.  In addition, some means to collectively (as a community) evaluate competing or new disruptive technologies, encourage commonalities, and guarantee maintenance of critical MMA codes would be another valuable service. Finally, there is an opportunity to improve practices in sustainable software development across the different communities that use different codes through an umbrella collaboration.  In particular, there is a need to foster
  integration across (software) ecosystems, and reducing the pigeonholing of
  software developers.  This can be partially accomplished by fostering workshops, conferences, or hackathons across these software ecosystems, but coordination in making this happen is sorely needed.
  
  \item \textbf{EM Followup Challenges:} Overall science output is increased, and redundancy reduced, when information is shared promptly and efficiently.  Tools for sharing information need to be developed or improved in several regimes.  These include searching localization regions, candidate follow-up, scheduling telescopes or instruments, observing plans, and data sharing. Standardized methods of information and data sharing also need development, together with incentives for opening access to data. The scheduling of telescope resources could be improved if they are done in an algorithmically-driven, adaptive manner that coordinates observations on multiple facilities and provides feedback to users. Responsive systems could make and distribute new plans in light of ongoing data acquisition and analyses of events.  Such scheduling of resources would both benefit time-critical transient events and long-term continual followup campaigns and can facilitate resource and data sharing.  These scheduling algorithms must also be swappable and extendable as the network of instruments grow, and will need to be implemented in a robust manner to ensure the necessary trust from facilities and the community.  
  \item \textbf{Storage, Archiving, Access, and Search of Data:} The major challenge of the data revolution is the volume of data that needs to be stored, archived, accessed, analyzed and interpreted. A community-driven body can lead the way in developing the cyberinfrastructure that allows for improved and simplified access to MMA data, respecting proprietary policies that applies to both data and descriptive metadata that vary greatly between facilities and experiments.  For instance, community broker services can facilitate searching for data in an area of a sky filtered by time and metadata, together with the creation of custom alerts for end users.  This service must be modularized with defined communications interfaces accessed via Application Programming Interfaces (APIs) to allow the underlying software to be swapped out, enhanced, or replaced, whilst ensuring that scientists can write their own clients if they wish. How this should be done and what descriptive metadata should be collected remain open questions and would interface well with work in the library/information sciences and computer science. Archiving issues include supporting the life-cycle of data such as providing archiving resources for data useful to the community.
  \item \textbf{Permissioning and MOUs: } A service that exposes subsets of data based on permissions, including derived data products, would be an important value added service for MMA. 
  Centralization of MOUs and permissioning would help enable coordination of follow-up, enable observatory facilities to optimize scheduling, as well as relative prioritization of events. It would also provide large facilities a centralized and standardized way of publishing and managing their data.
  Such a service would also promote the implementation of a cross collaboration authentication/authorization framework, 
  which creates opportunities for collaboration with researchers in computer science. 
  \item \textbf{Efficient and effective collaboration across disciplines: } 
  Ensuring a mutually rewarding collaboration between astrophysicists, astronomers, computer scientists and cyberinfrastructure specialists requires an understanding of the common areas of interest, the nature of the relevant expertise, and that scientists express their requirements so that designs can be \textit{targeted} at meeting those requirements. There is a clear need to facilitate collaboration between these groups together, establish a community-driven  set of requirements, record and disseminate decisions and designs, create funding streams for the production of cyberinfrastructure and tools, and verify that the work satisfies the requirements.
  \item \textbf{Education and Training:}  Hosting summer schools, developing a high quality set of interactive tutorials and notebooks, and enabling collaborative teamwork are major value-added activities to the MMA community, and are priorities for the larger astrophysics community. There is also a major need across astronomy and astrophysics to provide hooks into the deeper scientific and engineering knowledge, language, and cultural practices of the different major collaborations/facilities so that the subtleties of data analysis for different messengers are accounted for by external users. Establishing an unified approach to education on statistical techniques that will enable MMA analysis across the various messengers, and uphold a strong level of statistical analysis and results. Thoughtful design of these programs to be inclusive from the outset will also allow the MMA community to better reach under-represented groups as part of a larger effort. 
  \item \textbf{Collaboration Tools:} Services that enable collaborations, including wikis, mailing lists, scheduling and calendaring, video conferencing, centralized code repositories, data stores and management platforms all require a managing body. Such an agency could also provide the means to hold proprietary data in ``escrow'', or enable pre-registration of analyses to be executed on data as it becomes available, and can ensure that the APIs that enable communication between the diverse array of facilities are functioning as intended.  
  \item \textbf{Enabling Data Analysis Services:} Effective data analysis requires orchestration, which in turn necessitates the co-location of data and computational resources in a suitable analysis environment. Providing for such scalable frameworks for analysis of MMA data would be a major value added service to the community. One of the challenges is to quantify the requirements of data storage persistence, incoming data rate, processing latency, model robustness and update frequency, and prediction (service) latency. Another challenge is to corral the scalable compute systems (servers, compute clusters) for analyzing and serving data (centralized or on the cloud, on publicly-funded or commercial services) to provide services such as low-latency localization calculations based on incoming counterpart search data or large scale correlation or data analysis computations across the MMA spectrum. 
\end{itemize}

\bigskip
\noindent {\Large \bf An Institute for Multi-messenger Astrophysics}
\smallskip

There are no simple solutions to these problems; however, they provide significant opportunities for collaborative research between MMA and a broad range of computational disciplines. The success of MMA cyberinfrastructure will depend on both theoretical advances in how we represent data and the information they contain as well as the development of applications and systems that can scale to the complexity and size of data from a heterogeneous network of astrophysical facilities. Close interaction between MMA and programs within the NSF and other national agencies (DOE) such as the Office of Advanced Cyberinfrastructure (OAC), Cyber-Human Systems (CHS), and Information \& Intelligent Systems (IIS) can provide the necessary collaboration among data science, statistics, and astrophysics. 

The emerging issues of MMA cyberinfrastructure have been studied by a number of groups including an NSF-supported workshop at the University of Maryland in May 2018 \citep{2018arXiv180704780A} and SCiMMA. A key conclusion of these studies is the best way to meet the challenges is to establish a community-driven, decentralized institute for cyberinfrastructure for MMA. 

\textbf{We advocate support for a cyberinfrastructure institute for Multi-Messenger Astrophysics, to meet these community-driven challenges.} 
Such an institute must both lead and pragmatically respond to community opinion and needs.
The institute will have the technical capability to evaluate the success of
existing solutions, identifying new requirements from community needs, and
integrating existing or new cyberinfrastructure from other efforts. The institute will
ensure the stable delivery and maintenance of services, prioritize new
development according to its importance to scientific goals, and keep
abreast of technical developments in the academic and commercial sectors.
The institute will also integrate with other international infrastructures supporting
multi-messenger astronomy and work to support the required level of inter-operation.

The core mission of such an institute would be to provide value-added services and support to the emerging area of MMA without impeding the organic growth of the science.  As such, this institute would focus on cyberinfrastructure services that add to current capability without reproducing already existing solutions.  It will provide the glue between the different areas of MMA in a rational, flexible, and standardizable way.  Such an institute must be distributed in order to access the broad cross-section of researchers and expertise required to develop and support enhanced cyberinfrastructure critical to the success of MMA.

\pagebreak

\bibliographystyle{aasjournal}
\bibliography{references}
\end{document}